\documentclass[conference]{IEEEtran}
\IEEEoverridecommandlockouts
\usepackage{cite}
\usepackage{amsmath,amssymb,amsfonts}
\usepackage{algpseudocode}
\usepackage{algorithm2e}
\usepackage{graphicx}
\usepackage{textcomp}

\def\BibTeX{{\rm B\kern-.05em{\sc i\kern-.025em b}\kern-.08em
		T\kern-.1667em\lower.7ex\hbox{E}\kern-.125emX}}

\RestyleAlgo{ruled}
\begin{document}
	
	\title{Automation of Processor Verification Using Recurrent Neural Networks
	\thanks{This paper has been supported by Brno University of Technology (FIT-S-17-3994) and by the EC Horizon 2020 project MegaM@Rt2 ECSEL-JU, No. 737494.}
	}

	\author{\IEEEauthorblockN{Martin Fajcik, Pavel Smrz}
		\IEEEauthorblockA{\textit{Department of Computer Graphics and Multimedia} \\
			Brno University of Technology, Czech Republic \\
			\{ifajcik, smrz\}@fit.vutbr.cz}
		\and
		\IEEEauthorblockN{Marcela Zachariasova}
		\IEEEauthorblockA{\textit{Department of Computer Systems} \\
			Brno University of Technology, Czech Republic \\
			zachariasova@fit.vutbr.cz}
	}
	
	\maketitle
	
	\begin{abstract}
		
		When considering simulation-based verification of processors, the current trend is to generate stimuli using pseudo-random generators (PRGs), apply them to the processor inputs and monitor the achieved coverage of its functionality in order to determine verification completeness. Stimuli can have different forms, for example,  they can be represented by bit vectors applied to the input ports of the processor or by programs that are loaded directly into the program memory. In this paper, we propose a new technique dynamically altering constraints for PRG via recurrent neural network, which receives a coverage feedback from the simulation of design under verification. For the demonstration purposes we used processors provided by Codasip as their coverage state space is reasonably big and differs for various kinds of processors. Nevertheless, techniques presented in this paper are widely applicable. The results of experiments show that not only the coverage closure is achieved much sooner, but we are able to isolate a small set of stimuli with high coverage that can be used for running regression tests. 
	\end{abstract}
	
	
	\section{Introduction}\label{intro}
		
		
		
		With a raising demand on application-specific processors in today's market it becomes a necessity to come up with new improved approaches that tackle the challenging task of their complex verification. The current trend is to automate some of the routines of simulation-based verification, for example, generation of stimuli or targeting corner cases in coverage. 

		Pseudo-random stimuli generation is well defined in the SystemVerilog language standard~\cite{article:SVStandard:2013}, in the Universal Verification Methodology (UVM)~\cite{article:UVM:2015} and even in a new standard for portable stimuli (currently in its final revision state). Various pseudo-random stimuli generators (PRGs) can be utilized for that purpose, either those inbuilt in RTL simulators or external ones (e.g. in C++) connected through direct programming interface. Pseudo-randomness is achieved through constraints which are able to restrict the process of generation to gainful scenarios. This approach was devised to strike the balance between direct and random simulations. Instead of writing detailed verification patterns manually, the user provides a set of constraints and PRG generates a huge amount of stimuli in the order of seconds. 
		
		The completeness of stimuli generation can be measured by various metrics commonly known as coverage (we will consider RTL coverage metrics in this paper such as branch, statement, expression or functional coverage). We say that the processor reached coverage closure when the value of every monitored coverage metric is high enough to proclaim it as verified. Such feedback-controlled method based on coverage analysis is called Coverage-Driven Verification (CDV). A problem of this approach comes from redundancy inbuilt in randomness, because if the coverage feedback is not properly propagated to PRG and reflected by suitable constraints, the same redundant stimuli can be applied to Design Under Verification (DUV) without any coverage increase. Even intuitively we can observe that if the distribution of generated stimuli would reasonably change in every consequent stimuli generation (i.e. lower probability for already generated values), we would accelerate the verification processes and reach corner cases and coverage closure faster.
		
		From the practical point of view, this paper targets two industry-wide problems:
		\begin{enumerate}
			\item Speeding up the verification process by doing optimization of PRG constraints in such a way that coverage closure of DUV is achieved as fast as possible. 
			\item Seeking the smallest set of stimuli which reaches coverage closure for DUV. Such stimuli set is then ideal for regression testing.
		\end{enumerate}
		
		To resolve these problems, a new method for optimization of PRG constraints based on recurrent neural network (RNN) is proposed. This method has 3 significant contributions.
		Firstly, it automates CDV -- verification runs are repeated until either the coverage closure is reached or terminating conditions are met. Secondly, the proposed non-invasive solution can easily work with all verification environments that support PRG and coverage analysis. Thirdly, values of some RNN parameters and methods for their further tuning are suggested based on extensive experiments.
	
	\section{Related work}
		Combinatorial optimization techniques already appeared in works related to automation of verification processes. Kitchen and Kuehlmann in~\cite{inproceedings:KitchbenKuehlmann:2007} introduced a PRG using hybrid constraint solver based on Markov-chain Monte Carlo methods, which dynamically alters PRG distribution. However, these distributions were not biased from the coverage analysis like this work and further mentioned works are. In the past, two general approaches based on coverage analysis have been proposed: \textit{feedback-based CDV} (FBCDV) and \textit{CDV by construction} (CDVBC). FBCDV relies on a feedback from coverage analysis to be present during simulation and modifies the constraints to the PRG. In contrast, CDVBC approach relies on a generated external model of DUV which is used to generate stimuli designed to accurately hit the coverage tasks. CDVBC based approaches are commonly based on transformation of coverage tasks into Boolean logic (e.g. conjunctive normal form) and harvesting the power of SAT solvers \cite{inproceedings:HuangYeg:2010}\cite{article:ChengetAl:2014}. Although it is guaranteed that this systematic approach hits every coverage task, and using incremental SAT solvers it can even hit multiple tasks at once, it has been proven that the SAT problem is NP-complete and thus probably (if P!=NP) can never be optimal. 
		There are also feedback-based CDV approaches utilizing SAT solvers, e.g. in~\cite{inproceedings:HDLAnalysisApproach:2007} the DUV is reduced based on a current program slice (to reduce its complexity). PRG constraints are then treated as constraint-satisfaction problem and solved with a word-level SAT solver.
		The next category covers not fully automated approaches. For instance in~\cite{article:MarkovModelsApproach:2007}, the tool based on feedback-adjusted Markov Models called StressTest was proposed. The tool is capable of  on-the-fly optimization of constraints, but needs an assistance of engineer to provide a template describing interface protocols of the system.
		Finally, there are FBCDV approaches based on genetic algorithms (GA). Authors of~\cite{inproceedings:GoloubevaReordaViolante:2004} applied GA for automated generation of stimuli based on source code of specific software application. Naturally, such approach neglected all the details concerning the processor hardware irrelevant to the verified application. The most similar approach to this paper is proposed in~\cite{inproceedings:GACodasip:2015}. The work introduced capability of on-the-fly constraint optimization as well as seeking of an optimal stimuli set. GA was integrated directly into the UVM verification environment with profitable values of GA parameters. 
		The work presented in this paper builds on the GA approach (co-authored by M. Zachariasova). The GA optimizer has been replaced with a simple but powerful RNN. The proposed approach can be implemented with any verification environment that utilizes coverage analysis and PRG. Various neural network models with different network dynamics are proposed and compared to each other in Section \ref{sec:results}. We also provide a short comparison to the GA approach from~\cite{inproceedings:GACodasip:2015} in order to find out which kind of optimizer is more suitable for the CDV task. 
		
	\section{Proposed approach}
		\subsection{Integration of combinatorial optimization and verification environment} 
\begin{figure}[!h]
				\centerline{\includegraphics[scale=0.235]{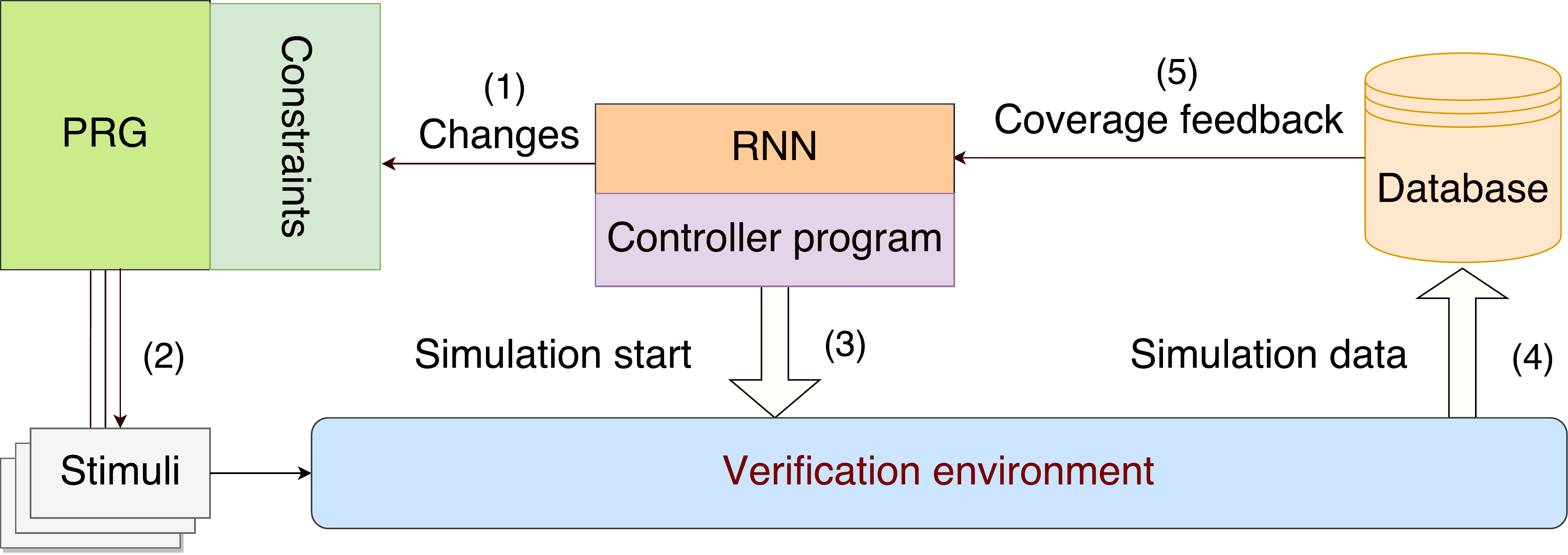}}
				\caption{Integration of RNN into the closed-loop solution.}
				\label{fig:RNNIntegration}
			\end{figure}

			Every optimization technique needs a method to evaluate quality of its solutions. This method is usually referred to as an objective function (OF) (also known as the cost or the fitness function). Since we are proposing a feedback-based CDV technique, the OF value needs to be determined based on the coverage analysis data. That is the reason why the proposed technique needs to be placed in a closed-loop architecture as shown in Fig.~\ref{fig:RNNIntegration}.
			
			Each loop in the automated verification life-cycle is composed of following steps:
			\begin{enumerate}
				\item The optimizer (in our case RNN) generates changes to PRG constraints. 
				\item A stimuli set is generated via PRG.
				\item The controller program starts a new simulation (or series of simulations) using the newly generated stimuli set. 
				\item At the end of simulation, most of HDL simulators are capable of exporting the database containing information collected during the simulation.
				\item A subset of the collected data (in our case various types of coverage metrics) is used to evaluate OF.
			\end{enumerate}
			Please note that this approach is non-invasive to the verification environment. 
			
			The PRG constraints define syntax and semantics of valid stimuli set. They can take many forms ranging from defining forbidden port combinations, valid instruction combinations, probabilities of generated values, to restrictions of already used variables. The constraints can also reduce generated stimuli to a subset valid for various portable stimuli scenarios.
			In our experimental evaluation we focus on verification of processors. Therefore, the optimizer is changing probability constraints on the instruction set level,the instruction level and the sub-instruction level (all these are closely explained in Fig.~\ref{subsubsection:acyclicgraphmodel}). The generated set of stimuli is in fact represented by a set of programs compiled into binaries. For the verification purposes we have used the UVM environment and all the scripts controlling RTL simulation, PRG and RNN were written in Python 2.7.
		
		\subsection{Hopfield model}
			Today, neural networks (NN) play an important role in the field of artificial intelligence, where they become an universal approach for modeling cognitive processes based on human brain image. They are currently a popular concept for the purpose of \textit{classification} and \textit{regression}. In \cite{article:Hopfield:1982}, J.J. Hopfield has shown that neural networks are also capable of dealing with \textit{combinatorial optimization} problems when, using his previously proposed RNN called Hopfield network (HN), he was able to solve traveling salesman problem. In this paper, we have been inspired by his idea and we propose a similar network models capable of dealing with automation of CDV.
			
\begin{figure}
				\centerline{\includegraphics[scale=0.11]{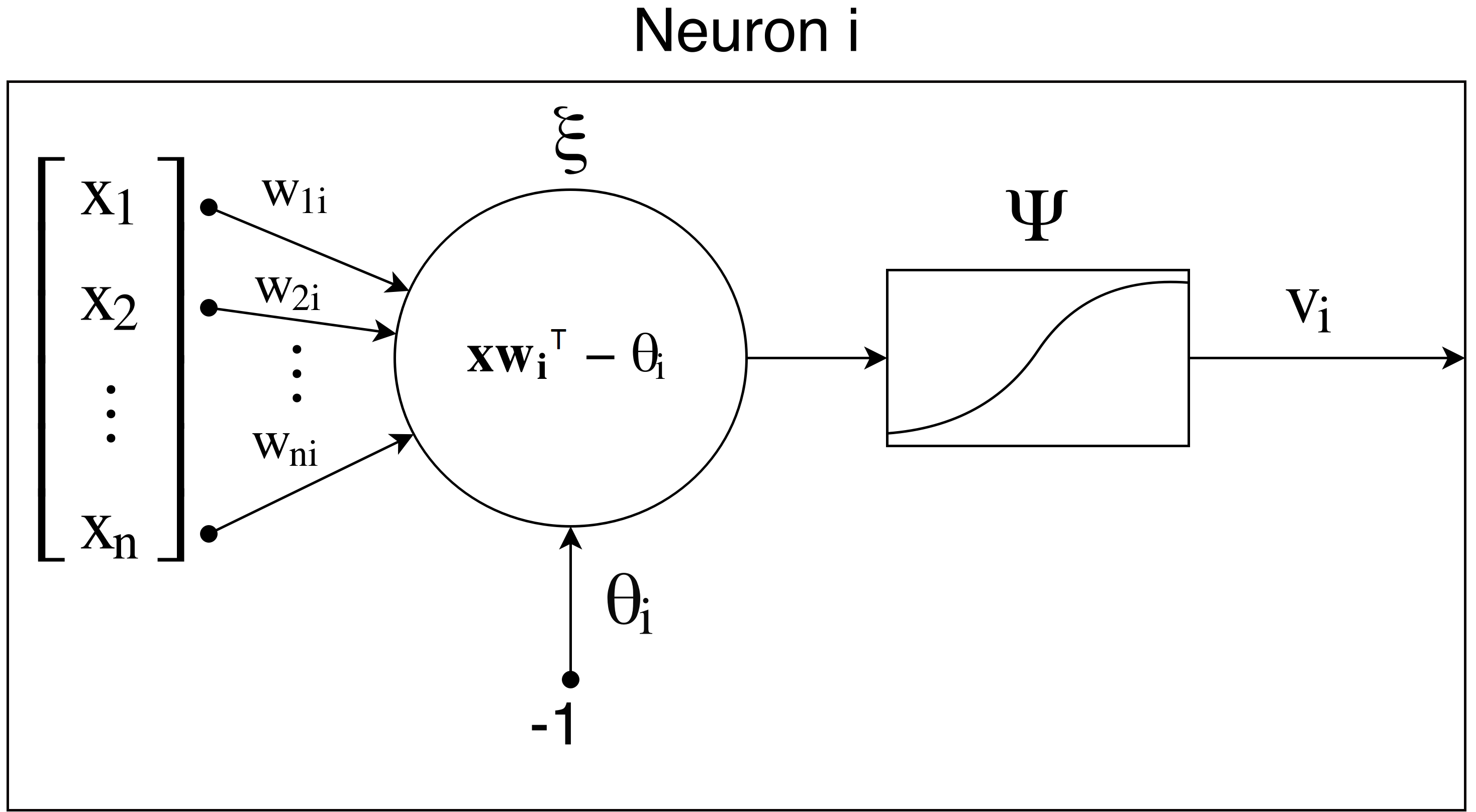}}
				\caption{Schema of the neuron used in Hopfield network.}
				\label{fig:neuron}
			\end{figure}

			HN work with autonomous units called neurons. We will label each neuron from the network with a number from arithmetic progression $m_{k+1}=m_{k}+1$ starting from the number $1$. A~neuron $i$ can be defined by a triplet: \textit{input vector} $\mathbf{x} = [x_1,...,x_n]^\intercal$ (also called \textit{feature vector}), output \textit{state} $v_{i}$, and \textit{threshold} $\theta_{i}$. Each neuron combination has a connection parameter called connection \textit{weight}. We will denote the value of such weight from the neuron $i$ to the neuron $j$ as $w_{ij}$. For HN of size $n$ weights between neurons can be defined by the weight matrix \eqref{eq:weightmatrix}.
			\begin{equation}
				\label{eq:weightmatrix}
				\mathbf{W} = [\mathbf{w_1},...,\mathbf{w_n}] = 
				\begin{bmatrix} 
				w_{11} & \cdots & w_{1n} \\
				\vdots & \ddots & \vdots \\
				w_{n1} & \cdots & w_{nn} 
				\end{bmatrix}
			\end{equation}
			The output of the neuron $i$ is given by the composition \eqref{eq:output} of linear basis function $\xi$ \eqref{eq:basef} with the sigmoid activation function (sometimes called the logistic function) $\Psi$ \eqref{eq:activationf}.
			
			\begin{equation}
				\label{eq:basef}
				\xi(\mathbf{x},\mathbf{W},\theta_{i}) = \mathbf{x}\mathbf{w_{i}}^{\intercal} - \theta_{i} = \sum_{k=1}^{n} w_{ki}x_{k} - \theta_{i}
			\end{equation}
			
			\begin{equation}
				\label{eq:activationf}
				\Psi(z) = \frac{1}{1+e^{-z}}
			\end{equation}
			
			\begin{equation}
				\label{eq:output}
				v_{i} = \Psi(\xi(\mathbf{x},\mathbf{W},\theta_{i}))
			\end{equation}
			The schema in Fig. \ref{fig:neuron} wraps up previous definitions. From the output $v_{i}$ of each neuron $i$ we can define a network output vector (or a state vector)  $\mathbf{v} = [v_1,...,v_n]^\intercal$.
			
			In addition, the network constructed from such neurons needs to satisfy several restrictions to meet the criteria of the HN. Let $N$ denote a set of NN neurons, then we define following constraints:
			
			\begin{enumerate}
				\item There must be a defined connection weight between each of two neurons (HN is a fully connected network).
				\item Diagonal values of $\mathbf{W}$ are equal to 0, so the value of the connection to the neuron itself is not used during the network evolution (see the further text for explaining the evolution process).
				\begin{equation}
				\label{eq:constraintZeroDiagonal}
				\forall i \in N : w_{ii}=0
				\end{equation}
				\item The network weights are symmetric.
				\begin{equation}
				\label{eq:symetricWeights}
				\forall i,j \in N : w_{ij}=w_{ji} 
				\end{equation}
			\end{enumerate}
			Since HN is a RNN, its state changes over time during its evolution. To be able to determine, when the network is in which state, we define the model time as follows: let a variable $t; t\in \mathbb{N} \wedge t>0$ be a time variable, then $v_i^{(t)}$ is a parametrized notation of the neuron outputs and $\textbf{v}^{(t)}$ is a network state which will be used to denote their value in certain evolution time $t$. Equation~\eqref{eq:output} that uses model time can now be rewritten as $v_{i}^{(t+1)} = \Psi(\xi(\mathbf{v}^{(t)},\mathbf{W},\theta_{i}))$. Next, we define a network trajectory in time $t$ to be a sequence of states $\textbf{v}^{(1)},\textbf{v}^{(2)}, ...,\textbf{v}^{(t)}$.
			There are two approaches used for the neuron activation policy. In the synchronous policy, all the units are activated simultaneously in each time step. In the asynchronous policy, only one unit at each time step is activated. The unit being activated can be chosen either randomly or sequentially. In the following text, we will assume that we are working with the random asynchronous policy.
			In order to solve optimization problems, the OF which is in context of neural networks commonly called the \textit{energy function} ($E$), needs to be defined. In~\cite{article:Hopfield:1982}, J.J. Hopfield defined a monotonically decreasing energy function, which can be used for simple problems like TSP, but does not seem to be applicable in this case (we will define our own in the further text). 

		\subsection{Proposed solutions}\label{propsolutions}
			It turns out that the mapping of the CDV problem to the RNN optimization model is quite straightforward. The state space of the processor is reachable via input stimuli with probabilities defined as constraints for PRG. Since probability $P(A)$ of an event $A$ has a domain of ${[0,1]}$, we can represent each of these probabilities with a neuron using a sigmoid activation. In \cite{misc:Lagoudakisneuralnetworks:1997}, an approach for designing custom energy function $E$ and tips for tuning $\textbf{W}$ and $\boldsymbol{\theta}$ are presented. As a result, we defined energy function as \eqref{eq:designedenergyfun}. Importance of each coverage metric $c \in C$ is expressed via relative weight coefficients $a_c$. Function $Coverage_c(\textbf{v})$ reflects the total coverage reached so far with stimuli generated from the network state $\textbf{v}^{(t)}$ together with all previously accepted NN states of evolution trajectory $\textbf{v}^{(1)}$,$\textbf{v}^{(2)}, ...,\textbf{v}^{(t-1)}$.
			
						
			\begin{equation}
				\label{eq:designedenergyfun}
				E(\textbf{v}^{(t)})= \sum_{c\in C}a_c P(Coverage_c(\textbf{v}^{(t)}))
			\end{equation}
			
			It can be seen that if the network evolution with trajectory $\textbf{v}^{(1)},\textbf{v}^{(2)}, ...$ starts in any state and only neurons leading to the energy difference $\Delta E\geq0;\Delta E = E(\textbf{v}^{(t)})-E(\textbf{v}^{(t+1)})$ are activated, the optimal solution state $\textbf{v}^*$ or a near optimal local minimum state will be reached eventually.
			 Note that  our objective is to reward new constraint settings which have not been used yet and lead verification process to so far unexplored processor states, instead of seeking the best initial configuration (as we would if we would not consider previously collected coverage data). Finally, $P$ represents a penalty function. In fact, it just transforms coverage values to fit the minimization problem expectations. Since the maximum coverage for each metric is 100\%, we define the penalty function as \eqref{eq:eqconstrain}.
			\begin{equation}
				\label{eq:eqconstrain}
				P(x)=(100-x)^2
			\end{equation}
			To clarify, we have been using various coverage metrics in the evaluation, e.g. functional (defined by covergroups and coverpoints in SystemVerilog), statement, branch, expression and FSM coverage (defined automatically by the simulation tool). For weight coefficients $a_c$, the value $1$ has been assigned to functional coverage (since it is the most important in terms of functional verification) and the value $0.0001$ to other coverage metrics.
			
			The next task is to select model parameters $\textbf{W}$, $\boldsymbol{\theta}$. These had to be chosen in such a way that reachability of any network state should be possible and probability of reaching these states should not be too small. 
			Ideally, during the NN evolution the output value of each neuron should be able to move in both directions, closer to $0$ or closer to $1$, but in general, all output values should have an average probability value $0.5$. This is the reason for defining a balanced weights model, which must satisfy the property \eqref{eq:balancedmodel}.
			\begin{equation}
				\label{eq:balancedmodel}
				\forall i \in N (v_{i}^{(t)}=0.5) \implies  \forall i \in N( v_{i}^{(t+1)}=0.5)
			\end{equation}
			Finally, the last modification is the sigmoid function \eqref{eq:activationf} which needs to be parametrized with the parameter $\lambda$ in order to adjust the function steepness as shown in \eqref{eq:activationfadjusted}. Intuitively, the slope needs to fit the network in such a way that wide range of values between ${[0,1]}$ can be selected. The value of this parameter was chosen experimentally, see Section~\ref{sec:results}.
			\begin{equation}
				\label{eq:activationfadjusted}
				\Psi(z)_\lambda= \frac{1}{1+e^{-z\lambda}}
			\end{equation}
			
			To finalize this section, the following two balanced weights models are proposed.
			\subsubsection{Bipolar model}
				The idea behind this concept is to split non-diagonal weights in $\textbf{W}$ into two equal groups, assigning one group with the value $1$ and other group with $-1$ in such a way that each neuron would have a half of the connections with the weight $1$, and the second half with the weight $-1$ as is shown in Algorithm \ref{alg:bipolarWeights}. In case of the odd number of non-zero weighted connections, weights of one group need to be adjusted in order to satisfy the balanced weights model constraint as is presented in Algorithm \ref{alg:modifyWeights}. It can be shown (by mathematical induction) that the weight matrix of this model cannot be symmetric.
				
				\begin{algorithm} \footnotesize
					\caption{ Assigning weights to the bipolar model\label{alg:bipolarWeights}}
					\KwData{Number of neurons $n$}
					\KwResult{Weight matrix \textbf{W} of size $n \times n$}
					$sign\leftarrow 1$\;
					\eIf{ $n-1$ is odd }{
						$oddModifier\leftarrow \frac{n-2}{n}$\;
					}{
						$oddModifier\leftarrow 1$;	
					}
					\For{$i\leftarrow 0$ \KwTo $n$}
					{
						$sum \leftarrow 0$\;
						\For{$j\leftarrow 0$ \KwTo $n$}
						{
							\uIf{ $i=j $}{
								$w_{i,j} \leftarrow 0$\;
							}{
								$sum \leftarrow sum+sign$\;	
								$w_{i,j} \leftarrow sign$\;
								$sign \leftarrow -sign$\;
							}
						}
						
						$\textsc{ModifyWeights}(oddModifier,sum,\textbf{W})$
					}
					\Return{\textbf{W}}
				\end{algorithm}	
				
				\begin{algorithm} \footnotesize		
				\caption{$\textsc{ModifyWeights}$(oddModifier,sum,\textbf{W})\label{alg:modifyWeights}}
				\If{$oddModifier \neq 1$}{
					\For{$j\leftarrow 0$ \KwTo $n$}
					{
						\tcc{Determine which group has more elements via sum}
						\If{\big($sum>0$ \textbf{and} $ w_{i,j}>0\big)$ \textbf{or}
							\big($sum<0$ \textbf{and} $ w_{i,j}<0\big)$ 
						}
						{
							$w_{i,j}=oddModifier *w_{i,j}$
						}
					}
				}
				\end{algorithm}
				
			\subsubsection{Acyclic graph model} \label{subsubsection:acyclicgraphmodel}
\begin{figure}
					\centerline{\includegraphics[scale=0.35]{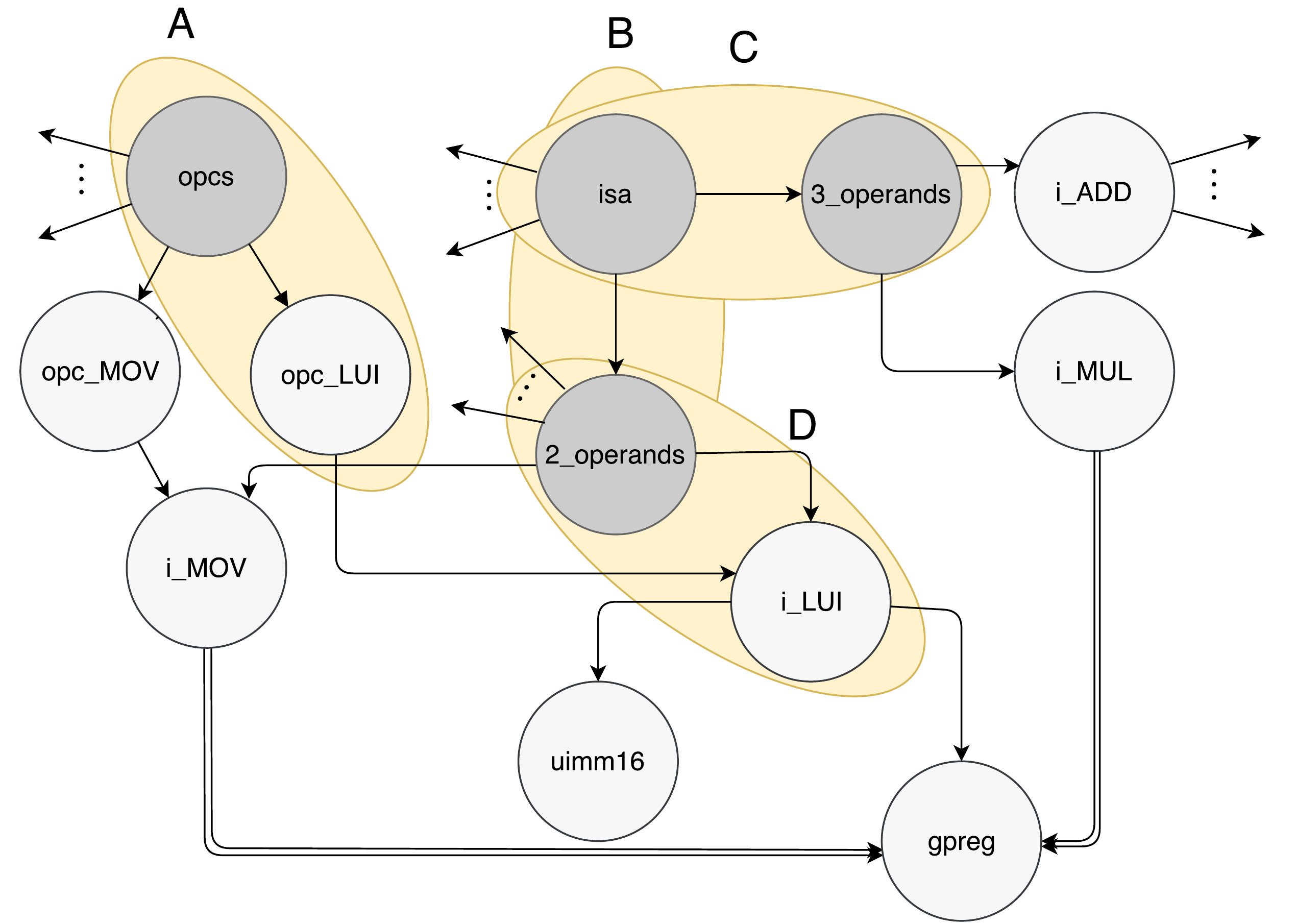}}
					\caption{A part of acyclic graph generated from ISA. $A$, $B$, $C$, $D$ are examples of unique instance vector members -- each represented by a neuron. Neuron combinations $(B,C)$ and $(B,D)$ have non-zero connection weights, since neurons $B$ and $C$ have the same set and the member of $B$ is a set of $D$. The probability of generating instruction \textit{i\_LUI} by PRG is equal to $P(isa)P(2\_operands|isa)P(i\_LUI|2\_operands)$.}
					\label{fig:ISAGraph}
				\end{figure}

				This model is designed specifically for the processor verification since it uses PRG generating programs. The NN topology comes from directed acyclic graph corresponding to the model of the instruction set architecture (ISA). ISA is usually designed on multiple levels. Sub-instruction level describes \textit{elements}, from which instructions are composed, for instance, instruction operation codes, registers or immediate operands. Instruction \textit{elements} are then put together from these sub-instructions. In addition, sometimes it is more beneficial to put elements or other \textit{sets} together into a \textit{set} and describe more complex elements on the instruction set level (e.g. all 2-operand instructions contained in \texttt{2\_operands} set have the same syntax, they just differ with opcodes). Our PRG constraint model corresponds to ISA model. Using such constraint model, we are able to set probability constraint for each member of the set, so each neuron represents a unique instance of the pair \textit{(\textit{set},\textit{member})}, where \textit{member} is either another \textit{set} or an \textit{element}. In this model, we have connected only those neuron instances in which the member of one instance was the set of another instance or both instances have the same set. For demonstration, see an example of such graph in Fig. \ref{fig:ISAGraph}. PRG using such constraint model generates instruction or sub-instruction with a probability of the connection to his parent with respect to probabilities the parent has with its grandparent and so on (analogically to a Bayesian network).
				The construction of $\textbf{W}$ matrix for this model is shown in Algorithm \ref{alg:acyclicweightdef}.
				\begin{algorithm} \footnotesize		
					\caption{Acyclic graph model weights\label{alg:acyclicweightdef}}
					\KwData{
						Number of neurons $n$,
						Unique instance vector $\textbf{c}$
					}
					\KwResult{Weight matrix \textbf{W} of size $n \times n$}
					$sign\leftarrow 1$\;
					\For{$i\leftarrow 0$ \KwTo $n$}
					{
						$n_{connections} \leftarrow 0$\;
						$sum \leftarrow 0$\;
						\For{$j\leftarrow 0$ \KwTo $n$}
						{
							\If{ $i=j $}{
								$w_{i,j} \leftarrow 0$\;
								$n_{connections} \leftarrow n_{connections}+1$\;
							}
							\ElseIf{$c_i.member = c_j.set$  \textbf{or} $c_i.set = c_j.member$ \textbf{or} $c_i.set = c_j.set$ }
							{	
								$w_{i,j} \leftarrow sign$\;
								$sum \leftarrow sum+sign$\;
								$sign \leftarrow -sign$\;
								$n_{connections} \leftarrow n_{connections}+1$\;
							}
						}
						
						\eIf{$ n_{connections}-1$ is odd }{
							$oddModifier\leftarrow \frac{n_{connections}-2}{n_{connections}}$\;
						}{
							$oddModifier\leftarrow 1$;	
						}			
						$\textsc{ModifyWeights}(oddModifier,sum,\textbf{W})$
						
					}
					\Return{\textbf{W}}
				\end{algorithm}
				
		
		\subsection{Model usage}
		These models are able to solve industry-wide problems presented in Section~\ref{intro}. Algorithm \ref{alg:continuousoptimization} can optimize PRG constraints on-the-fly and speed up the verification process, see results in Section~\ref{sec:results}. Algorithm \ref{alg:bestprograms} seeks the smallest processor program set that can be used as a regression test suite. 
		\begin{algorithm}\footnotesize	
			\caption{On-the-fly optimization of constraints\label{alg:continuousoptimization}}
			\KwResult{Coverage closure if iteration limit was not exceeded.}
			Initialize PRG and NN\;
			\If{there is some default PRG configuration}{
				Generate new stimuli set, run verification, get coverage feedback\;
			}
			Initialize network energy, from coverage feedback if there is any\;
			Initialize empty tabu list\;
			
			\For{time $t$ from $0$ to \texttt{settings.MAXtime}}
			{
				\If{tabu list contains each node}{
					Local minimum reached, return\;
				}
				Activate random neuron, apply new constraints and generate new stimuli\;
				Run verification above new stimuli set, get coverage feedback and calculate NN energy for this state\;
				\eIf{energy decreased \textbf{or} $\big($stochastic\_activation = true \textbf{and} $Random(0,1)< f_\epsilon(t) \big)$}{
					Activation is accepted, save new best energy energy and empty tabu list\;
				}
				{
					Activation is rejected, return latest activated neuron to its previous state and add it to tabu list\;
				}
			}		
			\vspace{-1mm}
		\end{algorithm}
		
		\begin{algorithm}\footnotesize
			\caption{Seeking the optimal stimuli set\label{alg:bestprograms}}
			\KwResult{Coverage closure and optimal stimuli set.}
			Initialize PRG, NN, initial energy, best program found...\;
			
			\If{there is some default PRG configuration}{	
				Run PRG \texttt{settings.BESTPROGRUNS} times, find best program, save it and  update energy\;
			}
			
			Initialize empty tabu list\;       
			\eIf{\texttt{settings.MAXEPOCHLENGTH}$>$ number of neurons in NN}{
				Epoch length = number of neurons in NN\;
			}
			{
				Epoch length = \texttt{settings.MAXEPOCHLENGTH}\;
			}
			\While{Coverage closure not reached and $t <$\texttt{settings.MAXtime} }{
				\For{Epoch length times}{
					Activate random neurons, find an activation with largest $\Delta E$ and save its program\;
				}
				Activate neuron leading to largest $\Delta E$\;
				Empty taby list and reinitialize variables\;
			}
				\vspace{-1mm}
		\end{algorithm}
	
	\section{Results}\label{sec:results}
		All experiments were realized with Intel Core i7 3610QM processor with 8GB RAM and 64-bit Debian 8.7. The objects of verification were two processors provided by Codasip, each of markedly different complexity. \textit{Codasip uRISC} (area 16k gates in 55LP, frequency 400MHz) is a 32-bit RISC micro-architecture with 4-stage pipeline, used mainly for demonstration and tutorial purposes. The NN controlling PRG for uRISC verification contained $41$ neurons. \textit{Codix Cobalt} (area 24k gates in 55LP, frequency 500MHz) is a high-performance production 32-bit RISC micro-architecture with 5-stage pipeline~\cite{misc:CodasipWebPage:2017}. The NN controlling PRG for Cobalt verification contained $1020$ neurons. All the generated programs were loaded directly into the memory of the processor, each program with an approximate length of 100 instructions. In one of the experiments we tried a range of random distribution models to initialize the initial NN state, such as triangular distribution and various uniform distributions with interval restrictions and it turns out that the uniform distribution with values drawn from the interval ${[0.4, 0.6]}$ seems to give the best results. After several experiments, the sigmoid steepness parameter $\lambda$ has been set to $0.9$. Finally, we have experimented with various values of the parameter defining the maximal length of the epoch and the value $20$ seems to be an ideal trade-off between the duration of seeking for an optimal stimuli set and its qualitative properties.
		
\begin{figure}
			\centerline{\includegraphics[scale=0.5]{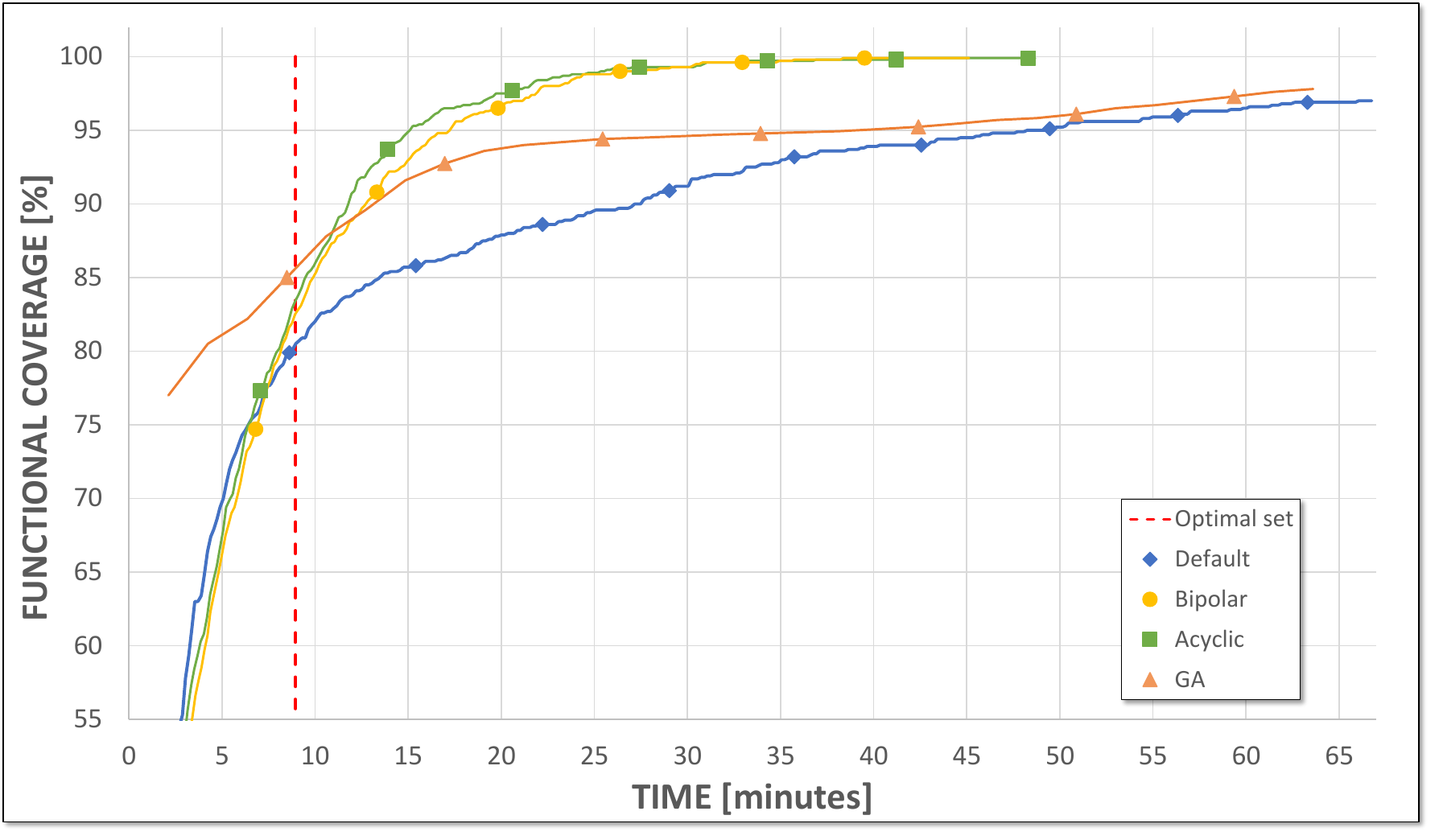}}
			\vspace{-1mm}
			\caption{Comparison of the proposed models and the GA during the Codasip uRISC processor verification. The optimal stimuli set was found in 12.05 hours, the final set was able to reach 100\% coverage in 8.94 minutes and contained 70 assembler programs. In contrast, default approach generated 390 programs in the shown interval. }
			\vspace{-2.5mm}
			\label{fig:UriscComparison}
		\end{figure}

\begin{figure}
			\centerline{\includegraphics[scale=0.5]{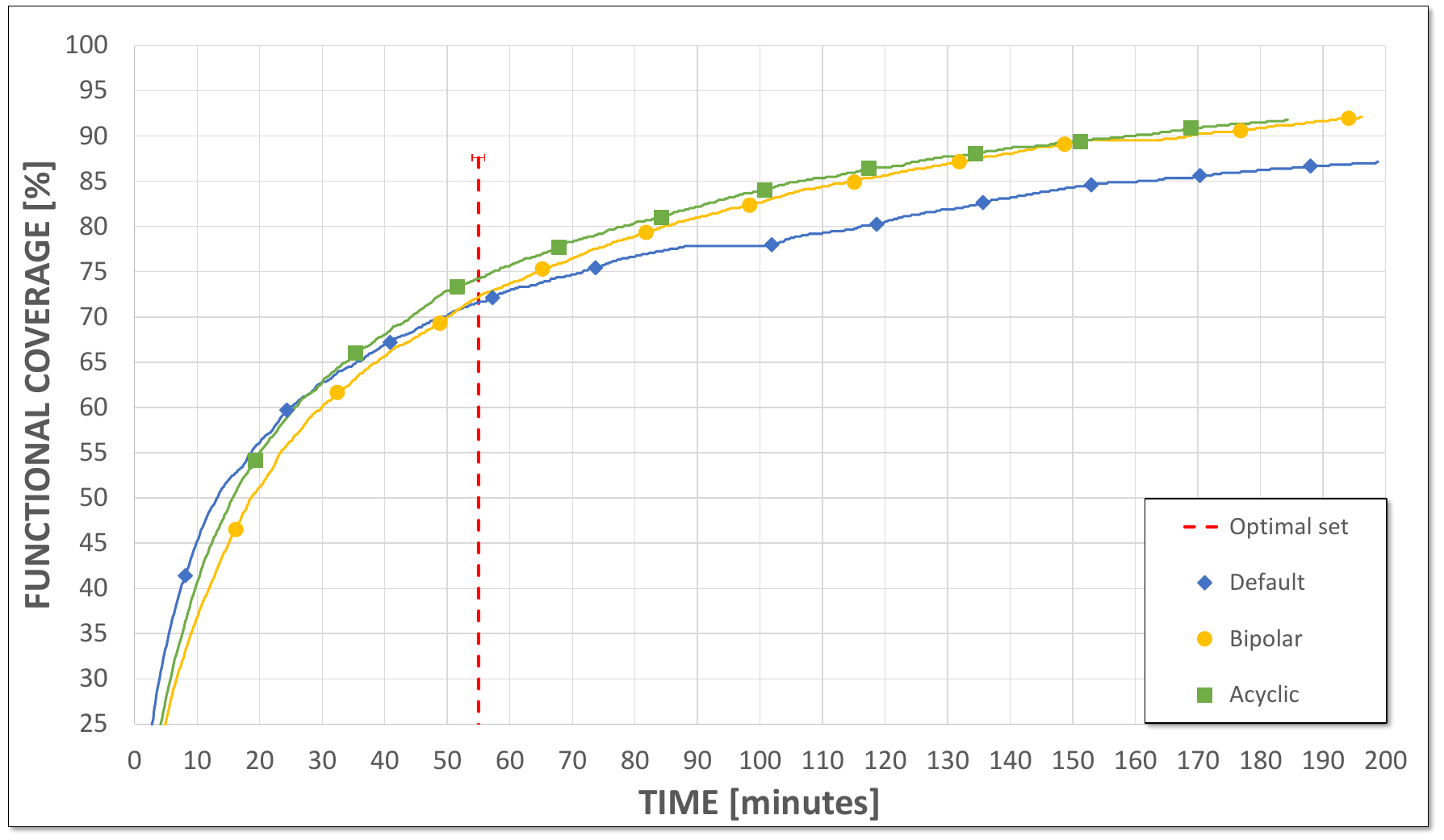}}
			\vspace{-1mm}
			\caption{Comparison of the proposed models during the Codix Cobalt processor verification. After 63.73 hours of the optimal stimuli set optimization the final set was able to reach 87.6\% coverage in 55.39 minutes and it contained 326 programs. The default approach generated 1000 programs during the measured interval.}
			\vspace{-3mm}
			\label{fig:CobaltComparison}
		\end{figure}

		The result of experiments for each model is shown in Fig.~\ref{fig:UriscComparison} and Fig.~\ref{fig:CobaltComparison}. Note that in our experiments, before the network started to modify constraints, each processor had a default PRG configuration available. The default function in both figures shows behavior with no involvement of the neural network or any other kind of coverage feedback-processing optimization algorithm. Fig.~\ref{fig:UriscComparison} also contains the comparison to the GA approach presented in~\cite{inproceedings:GACodasip:2015}. We do not have such comparison for the Cobalt processor since the GA results were not available. The vertical dash line displays time, in which the processor has been verified by optimal stimuli set. In addition, Fig. 6 shows the property of the balanced weight model, the average output value of the neuron is almost stable around the value $0.5$. 
	
		\begin{figure}
			\centerline{\includegraphics[scale=0.5]{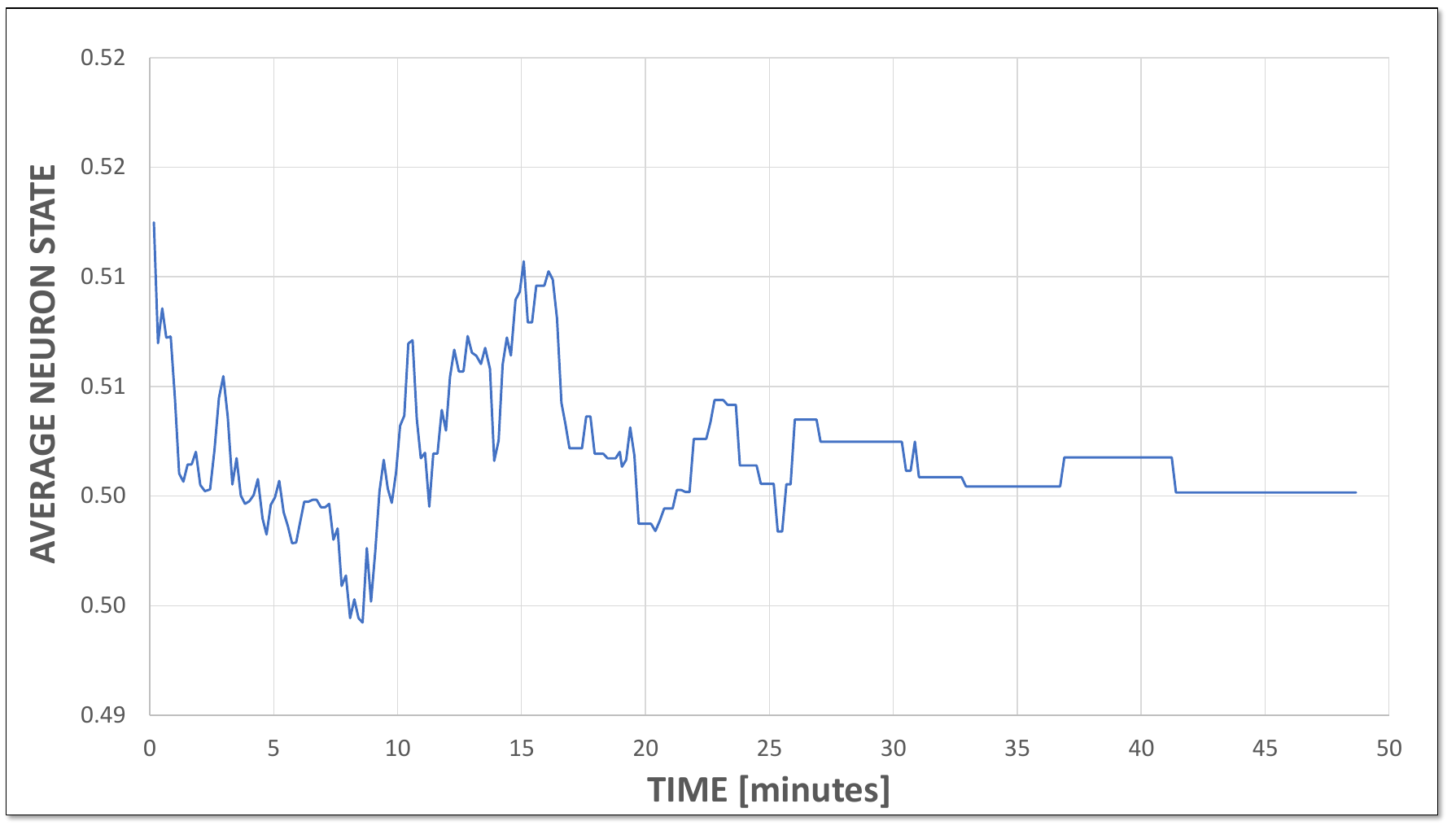}}
			\caption{The average output of network neurons during the Acyclic graph model experiment made on the Codasip uRISC.}
			\vspace{-3mm}
			\label{fig:BalancedWeightsModel}
		\end{figure}
	\section*{Conclusion}
	This paper has introduced a new approach of  automated feedback-based CDV accelerated by the RNN optimizer. Experiments have shown that the considered approaches achieved a large speed up of the verification process and have been able to find fairly small set of regression tests. However, in some initial time interval of the verification process, state-of-the-art approach (in our experiments referred to as 'default') seems to be more efficient than the proposed approaches. In the future work we would like to focus on this part of the proposed technique in order to develop further speed-ups and to define an interconnection of our work to a new portable stimuli standard after it will be publicly available. 
	
	\bibliographystyle{IEEEtran}
	\bibliography{references}
\end{document}